\documentstyle[twocolumn,aps,prb,psfig]{revtex}

 \renewcommand{\Re}{\mathrm Re}
 \renewcommand{\Im}{\mathrm Im}

 \title{
 On the theory of thermoelectric phenomena in
 superconductors}
 \author{Y.~M. Galperin$^{a,b}$,  V. L. Gurevich$^{b}$, V. I. Kozub$^{b}$,
 and A. L. Shelankov$^{b,c}$}
 \address{$^{(a)}$ Department of Physics,
 University of Oslo,  P. O. Box 1048 Blindern, N 0316 Oslo, Norway\\
 $^{(b)}$ Solid State Physics Division, A.~F.~Ioffe Institute,
 194021 Saint Petersburg, Russia\\
 $^{(c)}$ Department of Theoretical
 Physics,   Ume\aa\, University, 901 87 Ume\aa\, Sweden
 }
\date{March 20, 2001}
 \begin{document}
 \maketitle
 \begin{abstract}
 The theory of thermoelectric effects in superconductors is discussed
 in connection to the recent publication by Marinescu and Overhauser
 [\prb {\bf 55}, 11637 (1997)].  We argue that the charge
 non-conservation arguments by Marinescu and Overhauser do not require
 any revision of the Boltzmann transport equation in superconductors.
 We show that the charge current proportional to the gradient of the
 gap, $|\Delta |$ found by Marinescu and Overhauser, is incompatible
 with the time-reversal symmetry, and conclude that their
 ``electron-conserving transport theory'' is invalid.  Possible
 mechanisms responsible for the discrepancy between some experimental
 data and the theory by Galperin, Gurevich, and Kozub Pis'ma
 Zh. Eksp. Teor. Fiz. {\bf 17}, 689 (73) [JETP Lett. {\bf 17}, 476
 (1973)] are discussed.
 \end{abstract}
 \section{Introduction}\label{intro}

 The purpose of the present paper is to discuss some aspects of the
 kinetic approach to the thermoelectric properties of superconductors.
 As early as 1944 Ginzburg~\cite{Gi} suggested that in the presence of
 a temperature gradient, there appears in a superconductor a normal
 current of the form given by
 \[
 {\bf j}_n=-\alpha \nabla T \, .
 \]
 It was also pointed out by Ginzburg that the total current in the bulk
 of a homogeneous isotropic superconductor vanishes because the normal
 current is offset by a supercurrent ${\bf j}_s $ so that the total
 current in the bulk
 \[
  {\bf j}_n+{\bf j}_s=0 \, .
 \]
 This makes impossible the direct observation of the thermoelectric
 effect in a simply-connected homogeneous isotropic superconductor.
 Ginzburg considered also simply-connected anisotropic or inhomogeneous
 superconductors as systems where it is possible to observe
 thermoelectric phenomena by measuring the magnetic field produced by a
 temperature gradient.

 As indicated in Refs.~\onlinecite{GGK1,GGK2,GvH1} (see
 also~\cite{GGK3}), the best way to observe thermoelectric phenomena in
 superconductors, in particular, to measure the thermoelectric
 coefficient $\alpha $ is to make the superconductor a part of a
 bimetallic superconducting loop that may also contain weak links.

 Using the approach based on the Boltzmann equation for the normal
 excitations, the calculation of the coefficient $\alpha $ for
 impurity scattering has been made in
 \mbox{Refs.~\onlinecite{GGK1,GGK2}} -- see also
 reviews~\cite{AGGK1,AGGK2,Schoen81}. The expression for $\alpha$ has
 been later rederived in Ref.~\onlinecite{Ko} using the Green's function
 version of the nonequilibrium statistical operator approach. In that
 paper the role of paramagnetic impurities was also discussed. Based
 on the same method, an enhancement of the thermoelectric flux in
 superconductors containing nonmagnetic impurities with localized
 states near the Fermi energy was predicted in Ref.~\onlinecite{Ko0},
 see also Ref.~\onlinecite{Ko1}. As was proved in
 Refs.~\onlinecite{Ko,Ko0,Ko1}, the expressions for $\alpha$, obtained
 in Refs.~\onlinecite{GGK1,GGK2} for the case of nonmagnetic
 impurities remain valid for an arbitrary relation between the
 coherence length of the superconductor and the electron mean free
 path.

 Recently, Marinescu and Overhauser~\cite{MO} have proposed another
 method to calculate the transport coefficients $\alpha $. In their
 approach, the principal contribution to the thermoelectric effect in
 superconductors comes from the dependence of the superconducting gap
 $\Delta$ on the temperature. For some typical interval of temperatures
 and impurity concentrations their results differ from that of
 Refs.~\onlinecite{GGK1,GGK2} by several orders of magnitude.  Therefore, it
 is desirable to discuss the validity of their results.  In the present
 paper we compare these approaches.  We also briefly discuss  how
 the theoretical results are related to the existing experimental data.

 \section{Theory of Marinescu and Overhauser }\label{mar}

 Marinescu and Overhauser in Ref.~\onlinecite{MO} have proposed a method
 which they call ``electron-conserving transport equation''. They
 introduce distribution functions, $\tilde{g}_{{\bf k}\uparrow}$
 and $\tilde{g}_{{\bf -k}\downarrow}$ which differ from the
 distribution functions for the BCS excitations $f_{{\bf
 p}\uparrow,\downarrow }$ (below, the spin index is dropped).

 The nonequilibrium part of the distribution function [see Eq.~(47) of
 Ref.~\onlinecite{MO}] is
 \begin{equation}
 \delta g_{{\bf k}}=-{\hbar\tau_s\over m}\left[{\beta\epsilon_k^2f_k(1-f_k)
 \over TE_k}-{f_k\Delta\over E_k^2}\left({d\Delta\over dT}
 \right)\right]{\bf k}\cdot\nabla T.
 \label{4}
 \end{equation}
 Here, as in Ref.~\onlinecite{MO}, $\beta=1/k_BT$,
 $E_k=\sqrt{\Delta^2(T)+\epsilon_k^2}$, and
 $f_k=\left(e^{\displaystyle\beta E_k}+1\right)^{-1}$,  while
 $\epsilon_k=\hbar^2k^2/2m-\epsilon_F$ is the one-electron energy
 measured with respect to the Fermi level, $\epsilon_F$.

 The relaxation time $\tau_s$ is related to the relaxation time
 $\tau_n$ for impurity scattering in the normal state: For
 quasiparticle transitions from ${\bf k}$ to ${\bf k}'$,
 \begin{equation}
 \tau_s^{-1}=\tau_n^{-1}|E_k/\epsilon_k|(u_ku_{k'}-
 v_kv_{k'})^2\, .
 \label{5}
 \end{equation}
 For future convenience,  we write
 Eq.~(\ref{4}) as
 \begin{eqnarray}
 \delta g_{\bf k}&=& \delta g_{\bf k}^{(I)} +\delta g_{\bf k}^{(II)}
  \, , \label{ivb}\\
 \delta g_{\bf k}^{(I)}&=&-{\hbar\tau_s\over m}{\beta\epsilon_k^2
 \over TE_k}f_k(1-f_k)\, {\bf k}\cdot\nabla T
 \, ,
 \nonumber
 \\
 \delta g_{\bf k}^{(II)}&=&{\hbar\tau_s\over m}
 {f_k\Delta\over E_k^2}\left({d\Delta\over dT}
 \right){\bf k}\cdot\nabla T
 \; .
 \nonumber
 \end{eqnarray}
 Accordingly, the electric current density is split as
 $ 
 {\bf j}=
 {\bf j}^{(I)} +
 {\bf j}^{(II)}\, .
 $ 
 where
 \[
 {\bf j}^{(I)}= - \alpha^{(I)} \bbox{\nabla}T\, ,
 \qquad
 {\bf j}^{(II)}= - \alpha^{(II)} \bbox{\nabla}T 
 \]
 with
 \begin{eqnarray}
 \alpha^{(I)}
  & =  &\frac{4eN(0)}{3mk_BT^2}
 \int_{-\hbar\omega_D}^{\hbar\omega_D}\! \! \!d \epsilon\,
 \tau_s(\epsilon+\epsilon_F)
 \, f(1-f) {\epsilon ^{2}\over E}\, ,
 \nonumber \\
   \alpha^{(II)}   & =  & -
 \frac{4eN(0)}{3mk_BT^2}\int_{-\hbar\omega_D}^{\hbar\omega_D}\! \! \! \!d
 \epsilon \, \tau_s(\epsilon+\epsilon_F)
 {k_BT^2f\Delta\over E^2}
 \, {d \Delta \over dT} \, ,
 \nonumber
 \end{eqnarray}
with $N(0)$ being the density of states per spin.

 \section{Discussion}\label{dis}

Now we are in position to discuss the results of
Ref.~\onlinecite{MO}.  The first term, $\delta g_{\bf k}^{(I)}$ in
Eq.~(\ref{ivb}) comes from the coordinate dependence of the
temperature $T$ entering the Fermi-Dirac distribution function, and
therefore, is of true nonequilibrium origin. A nonequilibrium term
exists in the distribution function found from the Boltzmann equation
approach \cite{GGK1,GGK2} (cited as Eq.~(21) in Ref.~\onlinecite{MO})
with a very important difference: $\delta g_{\bf k}^{(I)}$ is {\it
even} under $\epsilon \rightarrow - \epsilon $ so that electrons
above and below the Fermi surface do not tend to compensate each
other as in Refs.~\onlinecite{GGK1,GGK2}. The opposite symmetry in
$\epsilon $ is the reason why the thermoelectric current obtained by
Marinescu and Overhauser is some five orders of magnitude larger than
that in Refs.~\onlinecite{GGK1,GGK2}.

 The origin of the second term, $g_{\bf k}^{(II)}$, is the
 $\bf r$-dependence of $\Delta = \Delta
 [(T({\bf r})]$ 
  as is obvious when $\delta g_{\bf k}^{(II)}$ is identically written
 in the following form:
 \[
 \delta g_{\bf k}^{(II)}=\tau_s
 f_k {\Delta({\bf r})\over E_k^2 }
 \left(
 {\bf
 v}\cdot
 {\partial\over{\partial {\bf r}}}
 \Delta ({\bf r})
 \right)
 \, , \quad
 {\bf v}= {\hbar {\bf k}\over m}\, .
 \]
 This means that in the approach of Ref.~\onlinecite{MO}, $\delta
 g_{\bf k}^{(II)}$ would exist irrespective of the origin of
 dependence of $\Delta({\bf r})$ on the coordinate $\bf r$.  For
 instance, such dependence may be due to the variation of the chemical
 composition of the superconductor or to the spatial variation of the
 strain.

 Even if the dependence $\Delta({\bf r})$ is due to one of these
 equilibrium mechanisms, the theory~\cite{MO} nevertheless predicts
 the current
 \begin{equation}
 {\bf j}= -\beta \; {\partial \Delta \over{\partial {\bf r}}}
 \label{jvb}
 \end{equation}
 where $\beta=\alpha^{(II)}(d \Delta/dT)^{-1}$.
 In our opinion,  such a current is forbidden.  Below we  give
 physical considerations supporting this statement.

 It is well-known that any linear response process can be classified
 as either reversible or irreversible (dissipative)~\cite{LanLif87}
 (see Ref.\onlinecite{Albert80} for a general discussion of
 nonequilibrium thermodynamics of superconductors, including
 thermoelectric phenomena).  The basis for the classification is the
 time reversal symmetry (T-symmetry). For instance, the charge current
 $\bf j$ induced by the electric field $\bf E$, in a normal conductor
 is irreversible. Indeed, the current changes its sign under time
 reversal whereas ${\bf E}$ remains intact, and the Ohm law, $\bf j=
 \sigma {\bf E}$, is not invariant relative to the $T$-symmetry
 transformation.  Thermoelectric current, ${\bf j}^{(T)}=-\alpha
 \bbox{\nabla}T$, is irreversible because the left-hand side changes
 its sign under time reversal whereas $\bbox{\nabla}T$ does not.
 Another example is the supercurrent, 
 $\bbox{j}_{s}= 2eN_{s} {\bf v}_{s}$, 
 proportional to the density of the Cooper pairs,
 $N_{s}$, and their velocity, ${\bf v}_{s}$. This relation is
 $T$-invariant since both current and velocity are $T$-odd quantities.
 Consequently, supercurrent is reversible and compatible with thermal
 equilibrium.

{}From this point of view, the current in Eq.~(\ref{jvb}) is
irreversible.  Indeed, the gap function $\Delta $ ($= |\Delta |$) is
unchanged by the time reversal transformation $\psi \rightarrow
\psi^{*}$, whereas the current changes its sign.  Unlike the
supercurrent, the irreversible current in Eq.~(\ref{jvb}) if existed
would be accompanied by a steady entropy production.  Being
incompatible with equilibrium, the current must be equal to zero.

\begin{figure}
  \centerline{\psfig{figure=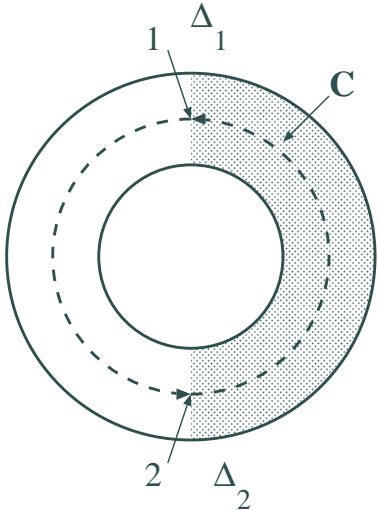,width=5cm}}
 \caption{On the equilibrium non-quantized magnetic flux. \label{fig1}}
 \end{figure}

The contradiction with the T-symmetry arguments can be also
demonstrated by the following \emph{gedanken} experiment. Consider a
ring built of two superconducting arms, left and right, both in
thermal equilibrium.  The arms are thick enough, so in their bulks the
magnetic field is completely screened.  The right arm is made of a
chemically inhomogeneous superconductor with a position-dependent gap,
$\Delta^{(r)}({\bf r})$, varying along the arm from $\Delta_1$ to $
\Delta_2$ if one moves counterclockwise, Fig.~\ref{fig1}.  The left arm
is built of a homogeneous superconductor with the gap
$\Delta^{(l)}$. The expression for the electric current, according to
Eq.~(\ref{jvb}) and the theory of Ref.~\onlinecite{MO}, reads
\begin{equation}
 {\bf j} =
 \left\{
 \begin{array}{lcr}
 \rule[-2.5ex]{0ex}{0ex}
 \frac{2e \hbar}{m}N^{(r)}_{s}
 \left[\bbox{\nabla}
 \chi^{(r)}
 -
 \frac{2\pi}{\Phi_0}
   {\bf A}
 \right] -
 \beta \bbox{\nabla}
 \Delta^{(r)}
 \, ,   &\text{right arm,}&\\
 \frac{2e \hbar}{m}N^{(l)}_{s}
 \left[\bbox{\nabla}\chi^{(l)}
  -
 \frac{2\pi}{\Phi_0}
   {\bf A}
 \right]\, ,& \text{left arm}.&
 \end{array}
 \right.
 \label{25b}
 \end{equation}
 Thus, there is an additional current  in the right arm given by
 Eq.~(\ref{jvb}). Since the current vanishes in the bulk of
 the ring,
 ${\bf j}=0$, from Eq.~(\ref{25b}) one gets the
 following equation for the phase $\chi $,
 \begin{eqnarray}
 (2\pi)^{-1} \bbox{\nabla}\chi^{(r)}&=&\Phi_0^{-1}{\bf A}+(\beta m/2e
 \hbar  N^{(r)}_{s})\bbox{\nabla} \Delta^{(r)}({\bf r})\, ,
 \nonumber \\
 (2\pi)^{-1} \bbox{\nabla}\chi^{(r)}&=&\Phi_0^{-1}{\bf A}\, . \label{35b}
 \end{eqnarray}
 Integrating the phase gradient along a contour $C$ in the bulk of the
 ring shown in Fig.~\ref{fig1} and remembering that the phase gets
an increment $2\pi n$ , one gets for the total flux in the ring,
 \begin{equation}
 \Phi = n \Phi_{0} + \Phi^{(MO)} \quad,\quad n = 0, \pm 1, \ldots
 \label{45b}
 \end{equation}
 where the additional flux $\Phi^{(MO)}$ is
 \begin{equation}
 \Phi^{(MO)} = \frac{\beta m \Phi_0}{2e \hbar}\int_{ \Delta_1}^{\Delta_2}
 \frac{d \Delta}{N^{(r)}_{s}(\Delta)} \, .
 \label{55b}
 \end{equation}
The lowest-energy state (at least for small values of $\Phi^{(MO)}$)
corresponds to $n=0$, and, therefore, the {\em equilibrium} flux of
the ring is predicted to be finite and equal to $\Phi^{(MO)}$.  
A similar conclusion was made in Ref.~\onlinecite{Oabst}.

Clearly, the finite magnetic flux and a finite electric current
generating the flux are not compatible with the underlying
time-reversal symmetry of the Hamiltonian of the superconductors.  It
is true that the state of the system may have symmetry lower that the
Hamiltonian: A ferromagnet gives an example of a system with
spontaneously broken T-symmetry. However, the underlying T-symmetry
guarantees the existence of two macroscopically allowed states related
to each other by the T-transform: for a magnet, theses are states with
reversed magnetizations. In other words, the underlying T-reversal
symmetry demands permission of any sign of the parameter that
quantifies the spontaneous symmetry violation.

In the case under consideration, the sign of the flux in the ring is
predetermined~\cite{Oabst}, given the geometry and the material
parameters.  This contradicts the time reversal symmetry, by which
the existence of  {\it two} equilibrium  states with the opposite
magnetic fluxes is compulsory.  Therefore, the current in
Eq.~(\ref{jvb}), which is the source of the spurious flux, must be
equal to zero.  (Incidentally, a finite $\Phi^{(MO)}$ would mean that
there are material-dependent corrections to the flux quantization
phenomena.)

These are our general arguments for why we think that the current
given by Eq.~(\ref{jvb}) should not exist.  Of course, the absence of
the equilibrium current which would be solely due to a spatial
dependence of $|\Delta |$ is well-known in the microscopic theory of
superconductivity \cite{deG89}.  Therefore, in our opinion, the
predictions of the theory~\cite{MO,Oabst} contradict the general
principles, and this is why we believe that the transport theory
suggested in Ref.~\onlinecite{MO} is invalid.

Although the derivation of the ``charge conserving transport
equation'' has not been presented by Marinescu and Overhauser
\cite{MO} in enough detail, let us try to specify those part in their
calculations which has lead to the above contradictions with general
principles.  In our opinion, it is mostly related to the transport
Eq.(39) of their paper.

The equation is formulated for the function $g_{{\bf k}\uparrow} $
which gives the occupation number of {\it bare} electrons [see
Eqs.~(37)-(38) of Ref.~\onlinecite{MO}].  The transport Eq.(39)
is identical to that which would be obtained by the standard procedure
based on the Liouville theorem~\cite{LanLif87}. As it is clear from
Eq.~(46), the Hamilton function $H({\bf r},{\bf p})$, which defines $
\hbar \dot{\bf k}= - \left(\partial H / \partial {\bf r}\right)$ and
$\dot{\bf r}= \left(\partial H / \partial \hbar {\bf k}\right)$, is
taken in Ref.~\onlinecite{MO} to be the BCS excitation energy 
$$H({\bf r}, \hbar {\bf k})=E_k({\bf r})=
\sqrt{\epsilon_{k}^{2}+|\Delta ({\bf r})|^{2}}\, .
$$ 
We believe that this procedure is qualitatively unsatisfactory: Due to
the electron-hole quantum coherence, the bare electrons are not good
semiclassical eigenstates, and the motion of wave packets built of
them cannot be described by the Hamilton equation, even approximately.

The main objection of Marinescu and Overhauser to the Boltzmann
equation approach \cite{AGGK1,AGGK2} is based on an apparent
non-conservation of ``bare electrons'' and thus of electric charge in
the course of the propagation of a wave packet.  As is well known, the
BCS Hamiltonian itself does not support detailed conservation of the
"bare electrons" while the charge conservation takes place only after
the quantum averaging over the quantum states provided the complex
pair potential $\Delta =|\Delta | e^{i \chi }$ satisfies the
self-consistency condition (see Appendix for the details). It has been
specifically emphasized in \cite{AGGK1,AGGK2} that the continuity
equation,\[
 e {\partial N\over{\partial t}} + \mathrm {div}\, {\bf j}=0
 \; ,
 \]
 bears no explicit relation to the Boltzmann equation.  In
 superconductors, the charge conservation law plays the role of a
 subsidiary equation allowing one to find the phase $\chi $ of the
 superconducting order parameter. Again, the conservation holds only as
 an average over different states and the self-consistency equation
 plays a crucial role to support the charge conservation.  Thus, the
 objections of Marinescu and Overhauser do not undermine the Boltzmann
 equation approach.  Furthermore, the same result for the transport
 coefficient was derived using the Green's function method.~\cite{Ko}

 \section{Experimental situation}\label{exp_dis}

 Let us briefly discuss experimental results.  The thermoelectric flux
 through a closed loop has been first measured by
 Zavaritskii~\cite{Zav}. The results exhibited no serious discrepancies
 with
 the theory of Refs.~\onlinecite{GGK1,GGK2}.
 However, in the later
 experiments~\cite{Pegrum1,Falco,Garland,vHG} the observed thermoelectric
 flux was much larger than the estimates from Refs.~\onlinecite{GGK1,GGK2}.

 In our opinion, the notable discrepancy between different
 experimental results as well between some of those results and the
 microscopic theory~\cite{GGK1,GGK2,Ko,Ko0,Ko1} remains a challenging
 problem. We present a brief review of different suggestions regarding
 this problem, see also p. 1856 of Ref.~\onlinecite{vHG} where a
 discussion of possible reasons for the discrepancy is presented.

 Two possible suggestions can be made regarding the source of puzzling
 discrepancies between the theory and the experimental
 results~\cite{Pegrum1,Falco,Garland,vHG}.  The first is focused on the
 differences between the realistic circuits used in
 Refs.~\onlinecite{Pegrum1,Falco,Garland,vHG} and the simple theoretical
 model~\cite{GGK1,GGK2}.

 The complications can arise, in particular, from the near-contact
 regions. In Ref.~\onlinecite{my2}, a thermoelectric loop consisting of an
 impure branch with higher $T_c$ and a pure branch with lower $T_c$
 (passive and active branches, respectively) was considered. It was
 shown that if (i) the electronic thermal conductance of the passive
 branch is much smaller than that of the active one, and (ii) there is
 finite thermal flux through the contact between these branches, then
 there exists a large contact thermoelectric contribution to the measured
 flux due to the phonon drag. The reason is that the phonon thermal
 flux in one of the materials can not be transformed to the electronic
 one abruptly at the contact. Rather, the transformation takes place
 within a near-contact layer of finite thickness where the phonon flux
 within the active branch is {\em much larger} than that in its bulk.
 As a result, the contact contribution can exceed the predictions of
 Refs.~\onlinecite{GGK1,GGK2} by a factor $\sim \epsilon_F/\Theta_D$
 where $\epsilon_F$ is the Fermi energy while $\Theta_D$ is the Debye
 energy. However this enhancement, although substantial, seems to be
 still too small to explain the magnitude of the effect reported in
 Refs.~\onlinecite{Garland,vHG}.

 Another suggestion is that the effects observed in
 Refs.~\onlinecite{Pegrum1,Falco,Garland,vHG} can be related to some
 temperature-dependent magnetic fluxes produced by external sources. A
 possible effect of such a sort was suggested in Ref.~\onlinecite{Pegrum} and
 later considered in detail in Ref.~\onlinecite{my1}.  This effect is related
 to a spatial redistribution of a background magnetic flux due to the
 temperature dependence of the London penetration depth, $\lambda$, of
 the superconductor.  Since the redistribution effect is proportional
 to $\lambda/L$, where $L$ is a size of the circuit, while the ``true''
 thermoelectric flux within the superconducting circuit is proportional
 to a smaller factor, $(\lambda/L)^2$, even very weak background fields
 can produce a temperature-dependent flux.
 Indeed, the effective
 near-surface area  $\sim \lambda L$ becomes rather large
  in the vicinity of $T_c$ where $\lambda > 1$ $\mu$m. Then, the
 non-screened magnetic field of the Earth, for instance, may generate a
 temperature-dependent magnetic flux as big as $\sim 10^3 \Phi_0$.

 It is worthwhile to note that if the diameter $d$ of the wires forming the
 loop is much less than $L$ the ``redistribution'' effect can be
 suppressed by a small factor $d/L$.  The reason is that the
 contributions of the ``inner'' and the ``outer'' parts of the wire to
 the redistribution effect have opposite signs. We believe that it is
 because of the suppression of the redistribution effect that the first
 observation of the thermoelectric flux~\cite{Zav} exhibited no serious
 discrepancies with the theory.

 To avoid the ``redistribution'' effect the authors of
 Refs.~\onlinecite{Garland,vHG} exploited the experimental set where the
 thermoelectric circuit had a shape of a hollow toroid, the
 thermoelectric flux being concentrated within its cavity.  The
 measuring coil was winded along the toroid.  The main idea was that
 the background fields were screened out by the bulk of the toroid.  In
 our opinion, there still existed a region between the coil and the
 bulk of the toroid where the background field can penetrate. In
 particular, this region included the near-surface layer of the
 thickness $\sim \lambda$. Thus the measured flux obviously included a
 contribution of this "outer" region.  Consequently, a temperature
 dependence of the penetration length brought about a
 temperature-dependent contribution provided that there existed some
 background field. The fact that, as reported in Refs.~\cite{Garland,vHG},
 the effect vanished for a homogeneous sample, in our opinion, cannot
 serve as proof of the absence of the redistribution effect. Indeed,
 if the measuring circuit is fully symmetric the contributions due to
 different parts of the toroid are compensated. However, for an
 inhomogeneous sample this symmetry will be absent, and the effect of
 the background field will be restored.

 We believe that to check experimentally the validity of the
 theoretical approach~\cite{GGK1,GGK2} it is practical, along with
 further studies of the thermoelectric flux under different geometries,
 to study thermoelectric effects of other type.  Among these effects,
 there is a specific interplay of a temperature gradient and a
 supercurrent in a superconducting film. Due to such an interplay a
 difference between the populations of the electron-like and hole-like
 branches of the quasiparticle spectrum is
 established~\cite{Falco1,Schmid,Schoen81,Shel}.  As a result, a difference
 $U_T$ between the electro-chemical potential and the partial chemical
 potential of the quasiparticles appears.  According to the
 experimental studies~\cite{Falco1,Clarke}, the measured values of
 $U_T$ agree with the theoretical predictions.  We would like to
 emphasize that the measurements of $U_T$ are {\it local} and
 consequently are much less sensitive to the above-mentioned masking
 redistribution effect.  Another way of local measurement is the
 thermoelectric modification of the Josephson effect in the SNS Josephson
 junction predicted in Ref.~\onlinecite{AG}.  The theoretical predictions
 obtained by the Boltzmann equation approach agree with the
 experimental results~\cite{RS}.

 \section{Conclusions}\label{con}

 In our opinion, the theory of thermoelectric effect in
 superconductors suggested by Marinescu and Overhauser in
 Ref.~\onlinecite{MO} is not valid.  In contradiction with general
 principles, the main contribution to the thermoelectric coefficient
 $\alpha$ calculated in Ref.~\onlinecite{MO} can be attributed to the
 spatial dependence of the order parameter $\Delta({\bf r})$, which is
 due to the temperature dependence of the gap $\Delta (T)$ when $T=
 T({\bf r})$.  As such, the spatial inhomogeneity of $\Delta $ does
 not disturb the thermal equilibrium.  We have shown that the
 existence of an equilibrium current $\propto \bbox{\nabla }\Delta $
 contradicts the time reversal symmetry, and, therefore, it cannot
 contribute to the thermoelectric coefficient.  In addition, being
 irreversible, the current found in \cite{MO} would lead to a steady
 entropy production in an inhomogeneous {\em equilibrium}
 superconductor, violating thermodynamics \cite{LanLif87,Albert80}.
 It is our belief that ``the electron-conserving transport equation''
 proposed in Ref.~\onlinecite{MO}, from which the above unphysical
 results are inferred, is erroneous.

 We have presented the conventional point of view,
 \cite{AGGK1,AGGK2,Galaiko} on the issue of the charge conservation in
 superconductors: It is an intrinsic feature of the BCS mean-field
 theory that the charge conservation may be violated {\it i.e.} ${\rm
 div} \bbox{j}_{n}\neq 0$, for any individual quasiparticle state
 $\psi_{n}$. However, the total electric current $\bbox{j}$, which is
 the sum over the quasiparticle states is {\em locally} conserved, {\it
 i.e.}  ${\rm div} \bbox{j}=0$, if and when the pair potential is
 self-consistent (see the Appendix for details).  Therefore, the Boltzmann
 equation supports the local charge conservation, and we disagree with
 the opposite claims in Ref.~\onlinecite{MO}.

 The results~\cite{GGK1,GGK2} of the Boltzmann equation approach are
 consistent with the experimental studies of the
 thermoelectrically-induced branch imbalance \cite{Clarke} and
 corrections to the critical current of the Josephson
 junction~\cite{RS}. The results of the thermoelectrically-induced
 magnetic flux through a close loop exhibit substantial
 scatter. Whereas the results \cite{Zav} are consistent with the theory
 \cite{GGK1,GGK2}, the giant thermoelectric flux observed in
 Refs.~\onlinecite{Pegrum1,Falco,Garland,vHG} is still not understood.
 We think that it is a challenging problem which is still open and it
 may require an account of additional sources of thermoelectrically-induced
 magnetic flux.  However, we are convinced that within the framework of
 the physical picture involving quasiparticle diffusion, the Boltzmann
 equation approach of Ref.~\onlinecite{GGK1} does not in principle
 require any revision.

 \acknowledgements

 This work has been partly done at the Centre of Advanced Studies,
 Oslo, Norway, and during the visits of two of the authors (VLG and
 VIK) to the University of Ume{\aa}, Sweden.  Supports by the Royal
 Swedish Academy of Sciences and by the Centre for Advanced Studies in
 Oslo are acknowledged.
 \appendix

 \section{Charge conservation}\label{charge}

 A quasiparticle in a superconductor is described by a two-component
 wave function $\bbox{\psi} = {u \choose v}$, with $u$ and $v$ being the
 electron and hole components of the quasiparticle, respectively.  The
 stationary wave function corresponding to the energy $E$ is found from
 the Bogoliubov - de Gennes equation~\cite{deG89}, $\hat{\cal H}\bbox
 {\psi}= E \bbox{\psi}$, with the matrix Hamiltonian
 \begin{equation}
 \hat{\cal H}=
 \left(
 \begin{array}{cc}
 \xi(\hat{\bf p } - {e\over c}{\bf A}) + U({\bf r})&
 \Delta  \\
 \Delta^{*} &
 -\xi(\hat{\bf p} + {e\over c}{\bf A})
 - U({\bf r})
 \end{array}\right).
 \label{txb}
 \end{equation}
 Here $\xi ({\bf p})= {\bf p}^2/2m- \epsilon_F $ For definiteness, we
 consider an isotropic s-wave superconductor so that the order
 parameter $\Delta ({\bf r})$ does not depend on the momentum.  Also,
 $\bf A$ is the vector potential, and $U$ is the potential energy,
 e.~g., due to impurities or the scalar electric potential.

 Generally, the eigenenergy $E$ in Eq.~(\ref{txb}) may be positive or
 negative. In the ground state (the condensate), the states with the
 negative energy are filled, and the positive-energy states are empty.
 As usual, the excitation is defined relative to the ground state, {\em
 i.~e.} it occupies an $E>0$ state or empties an $E<0$ state.  It is a
 property of Eq.~(\ref{txb}) that the eigenfunctions corresponding the
 energy $E$ and $-E$ are related to each other as~\cite{deG89} $${u
 \choose v}\quad\mbox{and}\quad{- v^{*}\choose \phantom{-}u^{*}}.$$
 This property allows one to express the contribution of the negative
 energy states via the positive energy ones, and, therefore, exclude
 the negative energies from consideration \cite{deG89,She97}.

We denote $$\psi_{n}={u_{n}({\bf r})\choose v_{n}({\bf r})}$$ as the wave
function of the excitation with the energy $E_{n}>0$ and $f_{n,\sigma
}$ as the distribution function of excitation ($\sigma =
\uparrow,\downarrow $ being the spin); here $n$ stands for the
quantum numbers other than spin. The observables can be expressed via
$u_{n},v_{n}$, and $f_{n, \sigma }$. \cite{deG89,She97}

The densities of charge, $Q({\bf r})$, and electric current, ${\bf
j}({\bf r})$, are given by the following expressions 
\begin{eqnarray}
Q & = & \sum\limits_{E_{n}>0}\left( 2e |v_{n}|^{2} + (f_{n \uparrow}
+ f_{n \downarrow })q_{n} \right)\, , \nonumber \\ {\bf j}& = &\sum
\limits_{E_{n}>0}
 \left(-1 + f_{n \uparrow}+ f_{n \downarrow}\right)
  \bbox{j}_{n}
               \label{yxb2}
 \end{eqnarray}
 where  the partial charge, $q_{n}({\bf r})$, and current, ${\bf
 j}_{n}({\bf r})$, densities are
 \begin{eqnarray*}
 e^{-1}q_{n}&=& |u_{n}|^{2}- |v_{n}|^{2}\, ,\\
 e^{-1}{\bf j}_{n}&=&
 \Re\,  \left(
  u^{*}_{n} \hat{\bf v} u_{n} -
 v^{*}_{n} \hat{\bf v}^{*} v_{n}
  \right)\, , \quad
 \hat{\bf v} = {1\over m}\left(\hat{\bf p} - {e\over c}{\bf A}
 \right)\, .
 \end{eqnarray*}
 Indeed, as it is rightly stated in Ref.~\onlinecite{MO}, the
 effective charge of the excitation $q_{n}$ is, generally, a function
 of the coordinate $\bf r$. Accordingly, the total charge of a wave
 packet built as a superposition of $\psi_{n}$'s varies while the
 packet propagates in an inhomogeneous superconductor. Of course, this
 violates the charge conservation on the level of an {\em individual}
 excitation. However, it is well-known that the local charge
 conservation is restored: (i) after summation over the states, i.~e.,
 one should consider only the total charge and current rather than
 $q_{n}$ and $\bbox{j}_{n}$; (ii) the pair potential $\Delta= |\Delta
 |e^{i \chi }$ is taken self-consistently rather than as an input.
 Below, we elaborate upon this point.

 Unlike the exact Hamiltonian of interacting particles, the BCS {\it
 effective} Hamiltonian does not commute with the particle number
 operator because of the presence of the anomalous average term
 $\Delta \hat{c}_{{\bf p}\uparrow}^{\dagger} \hat{c}_{-{\bf
 p}\downarrow}^{\dagger}+\mbox{h.\,c.} $.  For this reason, the charge
 conservation is not an automatic property of BCS-theory.  From
 Eq.~(\ref{txb}),
\begin{equation}
 \mathrm {div}\,  {\bf j}_{n} = - 4 \, \Im \, \Delta^{*} u_{n}
 v_{n}^{*}\, ,
 \label{zxb}
 \end{equation}
 so that lack of {\it detailed} current conservation, i.~e. $ {\mathrm
 div}\, {\bf j}_{n}\neq 0$, is obvious.

 For the total electric current density $\bf j$, one gets from
 Eqs.~(\ref{yxb2}) and (\ref{zxb}),
 \[
 \mathrm {div} \, {\bf j} = 4\, \Im \,  \Delta^{*}({\bf r}) F({\bf r})\, ,
 \]
 where $$ F({\bf r})= \sum\limits_{E_{n}>0}(1 - f_{n, \uparrow} - f_{n,
 \downarrow }) u_{n}({\bf r})v_{n}^{*}({\bf r}) .$$  As discussed in
 Ref.~\onlinecite{Galaiko}, the pair potential $\Delta $ ($\Delta^{*}$)
 serves as a source (sink) of the charge.  Again, the charge
 conservation is not guaranteed if the potentials in Eq.~(\ref{txb})
 are arbitrary inputs. However, the Gor'kov self-consistency
 condition demands that
 \[
 \Delta ({\bf r}) = g  F({\bf r})
 \]
 where $g$ is the coupling constant.  If the complex potential
 $\Delta$ is self-consistent, one readily sees  that $\mathrm {div}
 \, {\bf j}= 0$, {\em i.e.} the local charge conservation.

 These are our arguments against the point of view expressed by
 Marinesku and Overhauser that the Boltzmann equation scheme violates
 the local charge conservation.  To avoid confusion, another point
 should be mentioned. The lack of the detailed current conservation
 does not mean the absence of unitarity. Indeed, it generally follows
 from the Bogoliubov - de Gennes equation that the {\em quasiparticle}
 current $\bbox{j}^{(qp)}$,
 \[
 {\bf j}^{(qp)}_{n}=
 \Re \left(
  u^{*}_{n} \hat{\bf v} u_{n} +
 v^{*}_{n} \hat{\bf v}^{*} v_{n}
 \right)\; ,
 \]
 is a conserved quantity {\it i.e.} $\mathrm{div}\, {\bf
 j}^{(qp)}_{n}=0$, for any solution to the Bogoliubov - de Gennes
 equation.  The conservation of ${\bf j}^{(qp)}_{n}$ leads e.g. to the
 conservation of probabilities in the Andreev reflection problem.

 \end{document}